\documentclass[11pt]{amsart}
\usepackage{geometry}                
\usepackage{graphicx}
\usepackage{amssymb}
\usepackage{epstopdf}
\usepackage{amssymb}
\usepackage{yfonts}

\DeclareMathAlphabet{\mathpzc}{OT1}{pzc}{m}{it}
\usepackage{bm}
\usepackage{mathrsfs}
\usepackage{hyperref}
\usepackage[toc,page]{appendix}

\DeclareMathAlphabet{\mathpzc}{OT1}{pzc}{m}{it}

\newcommand {\CH}{\mathcal{H}}

\newcommand{\tg}{\textgoth{g}}

\title{Quantum theory from classical mechanics near equilibrium}

\author {  A. Schwarz\\ Department of Mathematics\\ 
University of 
California \\ Davis, CA 95616, USA,\\ schwarz @math.ucdavis.edu}

\begin{document}
\maketitle
    \begin{abstract} { We consider classical theories described by Hamiltonians $H(p,q)$  that have a non-degenerate minimum at the point where generalized momenta $p$ and generalized coordinates $q$ vanish. We assume that the sum of squares of generalized momenta and generalized coordinates is an integral of motion. In this situation,  in the neighborhood of the point $p=0, q=0$ quadratic part of a Hamiltonian plays a dominant role.  We suppose that a classical observer can observe only physical quantities corresponding to quadratic Hamiltonians and show that in this case, he should conclude that the laws of quantum theory govern his world.}
    \end {abstract}

    \section {Introduction}
    Recently I considered classical theories assuming that our devices can measure only some observables ("observable observables")\cite {NP}, \cite {SP}.
    I have shown that this consideration allows us to obtain  quantum theory  as well as numerous theories that go far beyond quantum mechanics
    (but are similar to conventional quantum mechanics in many ways). However, in my approach, the standard quantum theory was on equal footing 
    with its generalizations; it was not singled out in any way. In the present paper, I show that there is a very natural way to obtain quantum theory from classical 
    mechanics. This way can be regarded as a particular case of the general theory, but here it is considered independently without any reference to \cite {NP},\cite {SP}.
    (A short review of general theory and a derivation of the main result from it is given in Section 3.)

It seems that the results of the present paper can be used to realize experimentally quantum mechanics in the framework of classical mechanics.
    \section{Quantum from classical}
    
    Let us suppose that our world is described by classical theories with Hamiltonian functions $H(q,p)$ where $q=(q^1,...,q^n)$ are generalized coordinates and $p=(p_1,...,p_n)$ are generalized momenta with Poisson brackets $\{p_a,p_b\}=\{q^a,q^b\}=0, \{p_a,q^b\}=\delta_a^b$. (For simplicity, we work with systems having a finite number of degrees of freedom and assume that $H(q,p)$ is a polynomial function, but these restrictions are irrelevant.  Let us emphasize that from classical systems with a finite number of degrees of freedom, we get quantum systems with a finite-dimensional Hilbert space; to get conventional quantum mechanics, we need classical systems with an infinite number of degrees of freedom). We suppose that the Hamiltonian functions achieve a minimum at the point $p=0,q=0$ and that this point is a non-degenerate critical point. We assume that $\sum p_a^2+\sum (q^a)^2$ is an integral of motion for these Hamiltonians.
    
    It will be convenient to introduce complex coordinates $z_a=\frac {1}{\sqrt {2}}(q^a+ip_a)$  with Poisson brackets
 \begin{equation}\label {POI}     \{z_a,z_b\}=\{\bar z_a,\bar z_b\}=0,\{z_a,\bar z_b\}=-i\delta_{a,b}\end{equation}
    and regard the Hamiltonian functions and other observables as real functions of $\bar z,z$.

    We are interested in the case when the every term of  polynomials $H(\bar z, z)$ contains equal number of $\bar z$ and $z$; this means that $\langle z,z\rangle=\bar z_1 z_1+...\bar z_n z_n$ is an integral of motion or, in other words, $H$ is invariant with respect to transformations $z\to e^{i\alpha}z, \bar z\to e^{-i\alpha}\bar z$  where $\alpha$ is real. Then we can consider only observables $A(\bar z,z)$ that also have these properties. The most important role is played by observables \begin {equation}\label {CC}C^{ab}\bar z_az_b \end {equation}where $C^{ab}$ is a Hermitian matrix; only these observables are relevant in a small neighborhood of the point $z=0$ (higher order terms can be neglected). In Hilbert space $\CH$ with inner product $\langle z,v\rangle=\sum \bar z_iv_i$ these observables can be written in the form $\langle z,\hat Cz\rangle$ where $\hat C$ is a Hermitian operator. {\it We will assume that only these observables can be measured by devices available to the observer} (they are "observable observables").

    Two states of a classical system are equivalent for the observer iff measuring any "observable observable" in these states he obtains the same result. Representing states as
    probability distributions $\rho(\bar z,z),\rho' (\bar z,z)$ we can say that they are equivalent iff
    $$\int d\bar z dz C^{ab}\bar z_az_b \rho(\bar z,z) =\int d\bar z dz C^{ab}\bar z_az_b \rho'(\bar z,z)$$
    for all Hermitian matrices $C^{ab}$.
     
    Introducing notation   \begin{equation}\label {KK}K_{ab}=\int d\bar z dz \bar z_az_b \rho(\bar z,z) \end{equation}
    we can say that $\rho\sim \rho'$ iff $ K_{ab}=K'_{ab}.$ 
    
    One can say that the matrix $K_{ab}$ is obtained as an expectation value (an average) of the
    positive definite Hermitian matrix $\bar z_az_b$, hence it is also a positive definite Hermitian matrix.
    We see that the matrix $K_{ab}$ that describes the states of a classical system for our observer has all properties of density matrix in quantum mechanics except the normalization condition $Tr \hat K=1$. (We consider this matrix as a Hermitian operator $\hat K$ acting in Hilbert space $\CH$.)
    
    Let us check that that the observer will be forced to invent quantum mechanics to explain his experiments.
    
   
   It follows from the relation
   $$\int d\bar z dz C^{ab}\bar z_az_b \rho(\bar z,z) =
   C^{ab}K_{ab}=Tr \hat C\hat K$$
   that the right-hand side can be interpreted as an expectation value of observable $C^{ab}\bar z_az_b $ in the state $\hat K$ (again in agreement with quantum mechanics).
   
     In a neighborhood of the point  $z=0$, we can write

    $$H(\bar z, z)=h^{ab}
   \bar z_az_b+.....=\langle z,\hat h z\rangle+...$$
    where $...$ denotes higher order terms.
    
    The evolution of observables is governed by the Poisson bracket with $H$; for "observable observables", we can neglect higher-order terms in $H$; this leads to the standard quantum-mechanical relation
    \begin{equation}\label {H}
    i\frac {d\hat C}{dt}=[\hat C, \hat h].
    \end{equation}
Notice that in this relation $\hbar=1.$  
Instead of saying that we are working in small neighborhood of $z=0$, neglecting higher-order terms we can introduce new coordinates $u_a=\sqrt\epsilon z_a$ and work in these coordinates; the small parameter $\epsilon$ can be identified with
    Planck constant $\hbar.$

In conventional classical and quantum theories, we can consider either constant states and evolving observables or
evolving states and constant observables (in quantum mechanics, this is known
 as Heisenberg and  Schr\"odinger pictures). Here we can also use the relation
$Tr \hat C \hat K(t)=Tr \hat C(t)\hat K$ and (\ref {H}) to obtain the evolution equation for the state
\begin{equation}\label {K}
i\frac {d\hat K}{dt}=[\hat h,\hat K].
\end{equation}
One can get the same equation from (\ref{KK}) using the Liouville equation for $\rho$.

Notice that in equations (\ref{H}) and (\ref {K}) $\hat h$ can depend on time.

The appearance of probabilities follows from 
decoherence that can be proved as in \cite {ST},\cite {SP}. One should consider 
interaction with the environment modeled by  slow-varying (adiabatic) perturbation $\hat h(t)$. Let us denote by $\epsilon_n(t),\phi_n(t)$ eigenvalues and eigenfunctions of
$\hat h(t)$ for fixed $t$; we assume that all eigenvalues are distinct (the spectrum is simple). Then, in adiabatic approximation  (=neglecting time derivatives of $\phi_n(t)$), we can write the following equation  for matrix entries of
$\hat K(t)$  in the basis $\phi_n(t)$:
   \begin{equation}\label {AD}
    \frac {dK_{mn}}{dt}= i(\epsilon_m(t)-\epsilon_n(t))K_{mn}(t)
    \end {equation}
    It follows from this equation that non-diagonal entries
    ($m\neq n$) are unpredictable (they depend on the choice of adiabatic perturbation), but diagonal entries are predictable. Moreover, assuming that adiabatic perturbations are random and imposing some regularity conditions on the probability distribution, one can prove that the expectation value of non-diagonal entries vanishes (this fact is equivalent to the collapse of the wave function in the Copenhagen interpretation). 
    
  We proved that states in the neighborhood of equilibrium  are described by operators $\hat K$ that have all properties
  of density matrices  except the normalization condition $Tr \hat K=1$  and concluded that in the language of these operators one can prove decoherence and collapse of wave function.

The trace of $\hat K$ is an integral of motion of the equation (\ref {K}), hence we we can impose a condition $Tr\hat K=c$ where $c$ is some constant and solve the equation separately for every $c$.
If this condition is satisfied we can say that $\tilde K=c^{-1}\hat K$ has all properties of density matrix and  obeys
the Heisenberg equation
\begin{equation}\label{HEI}
i\hbar \frac {d \tilde K}{dt}=[\hat h,\tilde K].
\end{equation}
  with $\hbar=c^{-1}.$
 
  As we have seen after interaction with a random environment non-diagonal entries of the density matrix vanish.
       A diagonal density matrix can be considered as a mixture of  states $\bar z_az_b$ where $||z||=1$ (pure states) with weights equal to diagonal entries. If all eigenvalues are distinct  we obtain a unique decomposition of density matrix as a mixture of pure states and weights can be interpreted as probabilities.
    
    \section {Concluding remarks}
    Let us check that our statements can be derived from the general theory. In this theory, we consider a symplectic manifold (=phase space) or a Poisson manifold (a manifold with Poisson brackets of functions with standard properties). Probability distributions on the manifold are regarded as classical states. We denote this manifold by $M$. Real functions on $M$ are identified with classical observables. Let us fix a Lie algebra $\tg$ and a homomorphism  $\gamma\to a_{\gamma}(x)$  of this Lie algebra into the Lie algebra of functions on $M$ with respect to the Poisson bracket. Then one can define moment map $\mu$ of $M$ into the linear space $\tg ^*$ dual to $\tg$ sending the point $x\in M$ into
    $a_{\gamma}(x)$ regarded as linear functional of $\gamma$. This map can be extended to a map $\nu$ of probability distributions $\rho$ on $M$ (of classical states) into $\tg^*$ by linearity
    \begin {equation} \label {NU}\nu(\rho)=\int m(x)d\rho(x)=\int a_{\gamma}(x)d\rho(x).\end{equation}
    The right-hand side of (\ref {NU}) can be interpreted as
    the expectation value of the observable $a_{\gamma}$ in the state $\rho$. 
    
    Let us assume now that our devices can measure only observables $a_{\gamma}$ where $\gamma\in \tg$.
    It follows from (\ref{NU}) that in this case the observer can distinguish two states $\rho$ and $\rho'$ only if $\nu(\rho)\neq \nu(\rho')$. This means that from his viewpoint a state is described by the point $\nu(\rho)$ and the set of states is a convex envelope  $\mathcal N$  of the image of the moment map. It is shown in my papers how to develop a physical theory starting with the set of states (geometric approach).
    
    In  Section 2, we consider a space with coordinates $\bar z, z$ and Poisson brackets (\ref{POI}). We assumed that the observer can measure only observables given by the formula (\ref {CC}). They form a Lie algebra denoted by $\tg$. The moment map sends a point $(\bar z,z)$ into
    $\bar z_az_b\in\tg^*$. The set $\mathcal N$ consists of matrices $K_{ab}$.
    
    The observables (\ref {CC}) are invariant with respect to transformations $z\to e^{\alpha}z$ where $\alpha$ is real. If all observables share this property the observables (\ref {CC}) govern the behavior of our system near equilibrium. It follows that this behavior simulates quantum mechanics.
    
    One can express the main result of the present paper in the following way. Let us consider a quantum system in second quantization formalism, assuming that the number of particles is an integral of motion  (as in non-relativistic quantum mechanics). One can say that this system is obtained by the quantization of a classical system where creation and annihilation operators are regarded as canonically conjugate classical quantities. We assume that the energy of all states is non-negative. We argue that the dynamics of low-energy states of classical systems is governed by quantum mechanics.
    
    {\bf Acknowledgements} I am indebted to  L.  Alvarez-Gaume, A. Givental, D. Gross,  A. Kapusin,  A. Kitaev, 
    N.  Nekrasov, A. Vainshtein for useful discussions. This paper was written during my stay at Caltech and KITP; I am grateful to these institutions for their warm hospitality.

\end{document}